\documentclass[aps,prl,twocolumn,groupedaddress,showpacs]{revtex4}

\usepackage{latexsym}
\usepackage{dcolumn}
\usepackage[dvips]{graphicx}
\usepackage{amssymb}
\usepackage{graphics}
\usepackage{amsmath}
\usepackage{epsf}

\newcommand{\vk}{{\bf k}}

\begin{document}

%\twocolumn[
%\hsize\textwidth\columnwidth\hsize\csname@twocolumnfalse\endcsname
%\draft

\title{
Collective modes of the massless Dirac plasma}
% in thegraphene world} 
\author{S. Das Sarma and E. H.\ Hwang}
\affiliation{Condensed Matter Theory Center, Department of Physics,
  University of Maryland, College Park, 
Maryland  20742-4111 and Kavli Institute for Theoretical Physics,
Santa Barbara, California 93106} 
\date{\today}

\begin{abstract}
We develop a theory for the long-wavelength plasma oscillation of a
collection of charged massless Dirac particles in a solid, as occurring
for example in doped graphene layers, interacting via the
long-range Coulomb interaction. We find that the long-wavelength
plasmon frequency in such a doped massless Dirac plasma is explicitly
non-classical in all dimensions with the plasma
frequency being proportional to $1/\sqrt{\hbar}$. We also show that the
long wavelength plasma frequency of the $D$-dimensional superlattice
made from such a plasma does not agree with the corresponding $D+1$
dimensional bulk plasmon frequency. We compare and contrast such Dirac
plasmons with the well-studied regular palsmons in metals and doped
semiconductors which manifest the usual classical long wavelength
plasma oscillation.

\end{abstract}

\maketitle

A collection of charged particles (i.e. a plasma), electrons or holes
or ions, is characterized by a {\it collective mode} associated with
the self-sustaining in-phase density oscillations of all the particles
due to the restoring force arising from the long-range $1/r$ Coulomb
potential. The classical plasma frequency in three-dimensional (3D)
plasmas \cite{Jackson} is well-known to be $\omega_{3}=(4\pi
n_{3}e^2/m)^{1/2}$ 
where $e$ and $m$ are respectively the charge and the mass of each
particle, and $n_{3}$ is the 3D particle density. (In this paper, we
use $\omega_{D}$ and $n_{D}$ as the $D$-dimensional long-wavelength plasma
frequency and particle density respectively.)
%, whereas $V_d$ denotes the $d$-dimensional Coulomb interaction in
%the momentum space.)  
A solid state degenerate plasma \cite{Wolff,Pines,Mahan}
exists in metals and doped semiconductors
where free carriers can move around quantum mechanically in the ionic
lattice background. Such a degenerate quantum plasma has the
quantized version of exactly the same collective mode, the so-called
plasmon \cite{Wolff,Pines,Mahan}, which dominates the spectral weight
of the long-wavelength 
elementary excitation spectrum of an electron liquid. (We will use the
world `electron' generally throughout this paper to indicate either
electron or hole.) The
collective plasmon modes of solid state quantum plasmas have been
extensively studied  experimentally and theoretically over the
last sixty years in both metals and doped semiconductors. In the
present work, we study theoretically the collective plasmon mode in a
solid state plasma of massless Dirac fermions, as occurring for
example, in 2D graphene layers.
We define the Dirac plasma as a system of charged carriers whose
energy-momentum dispersion is linear, obeying the Dirac equation.

Our main qualitative result is that the massless Dirac
plasma is manifestly quantum, and does not have a classical limit in
the form of an $\hbar$-independent long-wavelength plasma frequency,
in a striking contrast to the corresponding parabolic dispersion
electron liquids familiar from the extensive study of plasmons in
metals and semiconductors \cite{Pines,Mahan,Ando}. 
The long-wavelength plasmon frequency of a Dirac plasma is necessarily
quantum with `$\hbar$' appearing manifestly in the long wavelength
plasma frequency in $D=1$, 2, 3 dimension (and in between).
By contrast the long
wavelength plasma frequency of ordinary electron liquids is
classical, and 
quantum effects show up only as nonlocal corrections in higher order
wave vector dispersion of the plasmon mode. 
This is quite unexpected
in view of the popular belief that the long wavelength quantum plasmon
dispersion is necessarily a classical plasma frequency
\cite{Wolff,Pines,Mahan}. The popular 
belief seems to be true for the usual parabolic energy dispersion, but
not for the linear Dirac spectrum.

We start from the fundamental many-body formula defining the
collective plasmon mode in an electron system:
\begin{equation}
\epsilon(q,\omega)=1-v(q)\Pi(q,\omega)=0,
\end{equation}
where $\epsilon(q,\omega)$ is the wave vector ($q$) and frequency
($\omega$) dependent dynamical dielectric function of the system,
with $\Pi(q,\omega)$ 
the irreducible polarizability and $v(q)$  the Coulomb interaction
between the electrons in the wave vector space. The zero of the
dielectric function in Eq.~(1) signifies a self-sustaining collective
mode, with the solution of Eq.~(1) giving the plasmon frequency as a
function of wave vector.
We first recapitulate the known results for the parabolic
dispersion electron system before discussing the novel collective
dispersion for massless Dirac plasma.

The Coulomb interaction in the wave vector space is given by the
appropriate $D$-dimensional Fourier transform of the Coulomb
interaction $v(r)=e^2/\kappa r$
\begin{subequations}
\begin{eqnarray}
v(q) & =&  \frac{4\pi e^2}{\kappa q^2}  \;\;\;\;\;\;\;\;\;\;\;\; D=3,  \\
     & =&  \frac{2\pi e^2}{\kappa q}  \;\;\;\;\;\;\;\;\;\;\;\; D=2, \\
     & =&  \frac{2 e^2}{\kappa} K_0(qa) \;\;\; D=1,
\end{eqnarray}
\end{subequations}
where we have introduced a background dielectric constant ($\kappa$) which, in
general, differs from unity in semiconductor based electron
systems, and $K_0$ is the zeroth-order modified Bessel function of the
second kind. We note that  $K_0(x) \sim |\ln(x)|$ for $x
\rightarrow 0$, and the length `$a$' in the 1D Coulomb
interaction in Eq.~(2c) characterizes the typical lateral confinement
size of the 1D electron system (ES) which is obviously necessary in
defining a 1DES. 

The irreducible polarizability function $\Pi(q,\omega)$ of an
interacting ES is, in general, unknown since
self-energy and vertex corrections cannot be calculated exactly.
A great simplification, however, occurs in
the long wavelength limit ($q \rightarrow 0$) when the dielectric
function, and consequently, the plasmon frequency is determined
entirely by the noninteracting irreducible polarizability,
the electron-hole `bubble' diagram. The noninteracting irreducible
polarizability is given by the expression: 
\begin{equation}
\Pi(q,\omega) = {g} \int \frac{d^Dk}{(2\pi)^D} \frac{n_F(\xi_k) -
  n_F(\xi_{k+q})} {\hbar \omega + \xi_k - \xi_{k+q}}F(k,q),
\end{equation}
where $\xi_k$ is the single particle energy dispersion i.e. $\xi_k =
\hbar^2k^2/2m$ for parabolic systems (and $\xi_k = \hbar v_F k$ for the
massless Dirac plasma), $n_F$ is the Fermi distribution function, and
$F(k,q)$ is 
the overlap form factor due to chirality.
For non-chiral systems $F(q,k)=1$. The factor
`$g$' in Eq.~(3) is the degeneracy factor: $g=g_sg_v$ where $g_s$ (=2)
is the spin degeneracy and $g_v$ is the valley or pseudospin degeneracy. 

Putting $\xi_k = \hbar^2k^2/2m$, we can easily calculate Eq.~(3) upto
the leading order in wave vector (i.e. the long wavelength limit) to
obtain
\begin{equation}
\Pi(q,\omega) \approx \frac{n_D}{m}\frac{q^2}{\omega^2}+
O(q^4/\omega^4).
\end{equation}
Combining Eqs.~(1)--(4) we immediately obtain the well-known
long-wavelength plasma frequency in a $D$-dimensional ES:
\begin{subequations}
\begin{eqnarray}
\omega_1^{(p)} &=& \sqrt{\frac{2e^2 n_1}{\kappa m}} q \sqrt{|\ln(qa)|} +
O(q^3), \\
\omega_2^{(p)} &=&\sqrt{\frac{2\pi n_2 e^2}{\kappa m}} q^{1/2} + O(q^{3/2}), \\
\omega_3^{(p)} &=& \sqrt{\frac{4\pi n_{3}e^2}{\kappa m}} + O(q^2), 
\end{eqnarray}
\end{subequations}
where $\omega_D^{(p)}$ denotes the long-wavelength ($q\rightarrow 0$)
plasmon mode in the $D$-dimensional parabolic dispersion ES
(with the carrier density $n_D$ per unit $D$-dimensional
volume) where the one particle energy is given by $\xi = \hbar^2
k^2/2m \rightarrow p^2/2m = mv^2/2$ classically (where $p=\hbar k$ is
momentum). The long wavelength plasmon frequencies for parabolic
dispersion systems given in Eq.~(5) are, of course, well-known and
have been verified experimentally extensively
\cite{Mahan,Ando,Li}. Our purpose of deriving 
Eq.~(5) is the explicit demonstration, to be contrasted below with the
corresponding massless Dirac plasma, that the long wavelength plasmon
frequency $\omega^{(p)}$ for parabolic systems is completely classical
since `$\hbar$' does not appear in the leading term of Eq.~(5) in any
dimension. (Note that the right hand side of Eq.~(5) has the explicit
dimensionality of time inverse, i.e. a frequency, in each dimension
since $n_D$ has the dimension of (length)$^{-D}$ in $D$-dimension.) The
second order dispersion correction term in Eq.~(5), i.e. the
$O(q^2,q^{3/2},q^{3})$ term in $D=3,$ 2, 1 respectively, is
fully quantum mechanical (i.e. `$\hbar$' shows up explicitly in the
non-local wave vector corrections), and is affected by interaction
corrections (both self energy and vertex corrections to the irreducible
polarizability).

Now we consider plasmons in the
$D$-dimensional massless Dirac plasma, where the single-particle energy
dispersion is linear, i.e. $\xi_k=\hbar v_F |\vk| \rightarrow vp$
classically, in $D=1,$ 2, 3. The long wavelength quantum plasmon
dispersion is still defined by the set of formula given by 
Eqs.~(1)--(3) with the explicit form of the noninteracting irreducible
polarizability being calculated with $\xi_k=\hbar v_F |\vk|$ in
Eq.~(3).

The long wavelength ($q\rightarrow 0$) form for the noninteracting
irreducible polarizability (Eq.~(3)) can be calculated for
linear energy dispersion relation (i.e. $\xi_k = \hbar
v_F k$) in all dimensions, giving
\begin{equation}
\Pi(q,\omega) =
\frac{g v_F k_F^{D-1}}{D(2\pi)^D} 
\frac{2\pi^{D/2}}{\Gamma(\frac{D}{2})}
  \frac{q^2}{\omega^2} + O(q^4/\omega^4),
\end{equation}
%\begin{subequations}
%\begin{eqnarray}
%\Pi(q,\omega) & = & \frac{g v_F}{2\pi} \frac{q^2}{\omega^2} +
%O(q^4/\omega^4)  \;\;\;\;\;\;\; D=1, \\
%\Pi(q,\omega) & = & \frac{g v_F k_F}{4\pi} \frac{q^2}{\omega^2} +
%O(q^4/\omega^4)  \;\;\; D=2, \\
%\Pi(q,\omega) & = & \frac{g v_F k_F^2}{6\pi^2} \frac{q^2}{\omega^2} +
%O(q^4/\omega^4)  \;\;\; D=3,
%\end{eqnarray}
%\end{subequations}
where $k_F$ is the Fermi momentum of the system and $\Gamma(x)$ is the
Gamma function. (We note that the
chirality factor $F(k,q)$ in Eq.~(3) does not influence the long
wavelength limit.)

Combining Eqs.~(1), (2), and (6), we get the following for the long
wavelength plasmon frequency, 
$\omega_D^{(l)}$, in $D=1$, 2, 
3 Dirac plasma:
\begin{subequations}
\begin{eqnarray}
\omega_1^{(l)}& =& \sqrt{r_s}\sqrt{\frac{g}{\pi}} v_F q
\sqrt{|\ln(qa)|} + O(q^3),  \\
\omega_2^{(l)} &=&  \sqrt{r_s} (g\pi n_2)^{1/4} v_F q^{1/2} +
O(q^{3/2}),  \\
\omega_3^{(l)} &=& \sqrt{r_s} \left ( \frac{32 \pi g}{3} \right
)^{\frac{1}{6}} n_{3}^{1/3}  v_F + O(q^2),
\end{eqnarray}
\end{subequations}
where we have introduced the dimensionless fine structure
constant $r_s(\equiv e^2/(\kappa \hbar v_F))$ for notational
simplicity.

Comparing Eqs.~(5) and (7) we see that $\omega^{(p)}$ and $\omega^{(l)}$ have
one important similarity and several striking differences. The
similarity is that the plasmon dispersion (i.e. the power law
dependence of the plasma frequency on wave vector) is the same in the
parabolic system and the massless Dirac plasma for all D. This is
indeed required under 
very general principles since for any Coulomb system, the long
wavelength plasmon dispersion is set by the continuity equation (or
equivalently, by particle conservation) to be
$\omega_D(q\rightarrow 0) \sim q^{(3-D)/2}$ as is obeyed by both
$\omega_D^{(p)}$ and $\omega_D^{(l)}$.

The most striking qualitative feature of $\omega_D^{(l)}$ in Eq.~(7), 
in sharp contrast with the usual $\omega_D^{(p)}$ in Eq.~(5), is that
`$\hbar$' appears explicitly in the {\it leading term}, not just the
subleading nonlocal corrections. A simple dimensional analysis of
Eq.~(7) shows that $\omega_D^{(l)} \sim O({\hbar}^{-1/2})$ in all dimensions
in contrast to the $O(\hbar^0)$ purely classical behavior of
$\omega_D^{(p)}$ in Eq.~(5). There is no
classical plasma frequency in the massless Dirac plasma, i.e. the
long-wavelength plasma frequency for ES with linear dispersion
explicitly depends on $\hbar$, and is
therefore, by definition, nonclassical. This absence of a classical
long wavelength plasma frequency in the Dirac plasma is a direct
manifestation of the relativistic Dirac nature of the underlying
quantum description, and such a Dirac plasma does not have a classical
plasma frequency.
Note that the nonclassical nature of the long wavelength plasma
oscillation of the Dirac plasma is independent of the chirality or
gaplessness of graphene, and arises primarily from the linear Dirac
spectrum.

Associated with the appearance of $1/\sqrt{\hbar}$ in the
long-wavelength plasma frequency of the Dirac plasma are several other
interesting properties distinguishing it from the standard parabolic
dispersion Schr\"{o}dinger plasma: (1) The density dependence of the
Dirac plasmon is different from the regular plasmon \cite{Hwang,Brey}
--- in particular, the density dependence is weaker in the sense that
$\omega_D^{(l)} \propto n^{1/3},\;n^{1/4},\;n^0$ in 
$D=3$, 2, 1 respectively in
contrast with $\omega_D^{(p)} \propto \sqrt{n}$ in all dimensions. In
general, the plasmon frequency in the Dirac plasma is given by
$\omega_D^{(l)} \propto n^{(D-1)/2D}$. (2) The
1D Dirac plasmon frequency is curiously density independent. (3) The
quantum coupling parameter (i.e. the effective fine structure
constant) shows up explicitly in the long-wavelength Dirac plasmon
frequency, $\omega_D^{(l)} \propto \sqrt{r_s}$.
(4) The long wavelength Dirac plasmon $\omega_D^{(l)}$ goes as
$\hbar^{-1/2}$ for all dimensions whereas the long wavelength regular
plasmon $\omega_D^{(p)}$ goes as $m^{-1/2}$ in all dimensions.

Before concluding, we consider another interesting and peculiar feature
of the Dirac plasmon distinguishing it from the regular plasmon. We
consider collective modes of periodic arrays of 2D Dirac plasma layers
(for example, a graphene superlattice made of parallel 2D graphene sheets
in the direction transverse to the 2D graphene plane) and of 1D
Dirac plasma nanoribbons (i.e. a graphene superlattice made of
identical 1D graphene nanoribbons placed parallel to each other in the
2D plane). Collective plasmon modes of such 2D \cite{dassarma} and 1D
\cite{Lai} superlattices
have been theoretically studied in the context of regular parabolic
systems, and have been experimentally observed in doped GaAs
multi-quantum well and multi-quantum wire structures.

The main physics to be considered in describing the collective plasmon
modes of such superlattices is the inclusion of the inter-layer or
inter-ribbon Coulomb interaction, which will necessarily couple all
the layers (or the ribbons) due to the long range nature of the Coulomb
potential. This changes the fundamental collective mode equation
(Eq~(1)) to an infinite matrix equation:
\begin{equation}
\left | \delta_{ll'}-v_{ll'}(q,\omega) \Pi_{l'}(q,\omega) \right | =0,
\end{equation}
where $\Pi_l=\Pi$ is the irreducible polarizability of each 2D layer
(or 1D ribbon), which is exactly the same 
polarizability considered in Eqs.~(3) and (4).
In Eq.~(8), $v_{ll'}$ is the Coulomb interaction
between the $l$ and the $l'$ layer or ribbon in the periodic array,
which is given by
\begin{subequations}
\begin{eqnarray}
v_{ll'} & = &\frac{2\pi e^2}{\kappa q} e^{-qd|l-l'|} \;\;\; D=2, \\
v_{ll'} & = &\frac{2e^2}{\kappa}\left [ K_0(qa) + K_0(qd|l-l'|) \right ]
\;\;\; D=1,
\end{eqnarray}
\end{subequations}
where `$d$' is the superlattice period (to be distinguished from the
length `$a$' in $D=1$ which defines the lateral width of each ribbon).

The periodic invariance of the superlattice and the associated Bloch's
theorem allow an immediate solution of the infinite-dimensional
determinantal equation defined by Eq.~(8), leading to the following
collective plasmon bands for the superlattice structure:
\begin{subequations}
\begin{eqnarray}
\tilde{\omega}_{2s}(q;k)& = &\omega_2(q) S_2(q,k) \;\;\; D=2, \\
\tilde{\omega}_{1s}(q;k)& = &\omega_1(q)S_1(q,k) \;\;\; D=1,
\end{eqnarray}
\end{subequations}
where $\tilde{\omega}_{Ds}$ is the plasmon band frequency for the superlattice
($D=2$ for the multilayer and $D=1$ for the multiribbon periodic
arrays) and $\omega_D$ is the corresponding 2D ($D=2$) and 1D ($D=1$)
plasmon modes discussed in Eqs.~(5) and (7). The wave vector $q$ in
Eq.~(10) is the same conserved 2D or 1D wave vector in each individual
2D layer or 1D nanoribbon defining the plasmon dispersion relation
$\omega_D(q)$ whereas the additional wave vector $k$ is a new continuous
parameter defining the superlattice plasmon band (arising from the
periodicity in the array structure). The band wave vector $k$ is
restricted to the first superlattice Brillouin zone, $k\le
\pi/d$, in the reduced zone scheme. For the 2D layer superlattice, if
each layer is assumed to lie in the $x$-$y$ plane, then $k=q_z$ is
along the superlattice direction of the $z$-axis. For the 1D ribbon
superlattice, if each ribbon is assumed to be along the $x$-axis
(i.e. $q=q_x$) with a width of {\it `a'} defining the ribbon in the
$y$-direction, then $k=q_y$ is along the superlattice direction of the
$y$-axis.

The function $S_D$ in Eq.~(10) is a form factor arising from the
Coulomb coupling between all the layers and the ribbons forming the
periodic array, and is given by 
\begin{subequations}
\begin{eqnarray}
%\begin{array}{ll}
S_2 &=& \sum_{l'}e^{-q|l-l'|d- i q_z|l-l'|d} \;\; D=2, \\
S_1&=&\sum_{l'}\left [ K_0(q|l-l'|d)\cos(lq_yd)+K_0(qa) \right ]\;
D=1. \;\;\;\;\;\;\;\;\;\; 
%\end{array}
\end{eqnarray}
\end{subequations}
Combining the above equations for superlattice plasmons, we get the
following long wavelength ($q\rightarrow 0$) plasmon bands for 2D and
1D arrays in parabolic and Dirac plasma systems, respectively:
\begin{subequations}
\begin{eqnarray}
\tilde{\omega}_{2s}^{(p,l)}({\bf q}) & = &\omega_2^{(p,l)}(q) \left [
  \frac{\sinh(qd)}{\cosh(qd)-\cos(q_zd)} 
\right ]^{1/2},  \\
\tilde{\omega}_{1s}^{(p,l)}({\bf q}) & = &\omega_1^{(p,l)}(q)
\nonumber \\
&\times&\left [
  K_0(qa)+2\sum_{n=1}^{\infty}K_0(nqd)\cos(q_ynd) \right ]^{1/2}.
\;\;\;\;\;\;\;\;\;
\end{eqnarray}
\end{subequations}
Eq.~(12) above defines plasmon bands for superlattice arrays
made out of periodic 2D layers and 1D ribbons in parabolic and linear
plasma systems.

An interesting quantum feature of $\tilde{\omega}_{Ds}^{(l)}(q,k)$ is
apparent when one looks 
at the long-wavelength plasmon ($q\rightarrow 0$) at the band edge
$k=0$, and compares $\tilde{\omega}_{Ds}^{(l)}(q,k=0)$
with $\tilde{\omega}_{Ds}^{(p)}(q,k=0)$. We get 
\begin{subequations}
\begin{eqnarray}
\tilde{\omega}_{2s}^{(p)}(q;q_z=0) & = & \left ( \frac{4\pi
    \tilde{n}_3 e^2}{\kappa 
    m}\right )^{1/2} \; {\rm with} \; \tilde{n}_3=\frac{n_2}{d}, \;\; \\
\tilde{\omega}_{1s}^{(p)}(q;q_y=0) & = & \left ( \frac{2\pi
    \tilde{n}_2e^2 q}{\kappa m} 
\right )^{1/2} {\rm with} \;  \tilde{n}_2=\frac{n_1}{d}, \;\;\;\;\;
\end{eqnarray}
\end{subequations}
and
\begin{subequations}
\begin{eqnarray}
\tilde{\omega}_{2s}^{(l)}(q;q_z=0) & = & \sqrt{r_s}(4\pi g)^{1/4} \left (
  \frac{\tilde{n}_3}{d} \right 
)^{1/4} v_F, \\
\tilde{\omega}_{1s}^{(l)}(q;q_y=0) & = & \sqrt{r_s} \sqrt{\frac{g}{d}}
v_F \sqrt{q}. 
\end{eqnarray}
\end{subequations}
We note that Eq.~(13) for the usual parabolic electron plasma has the
appropriate physical limit at the band-edge $k=0$, where the
$D$-dimensional superlattice plasmon should have the precise character
of the corresponding $(D+1)$-dimensional bulk plasmon in the long wavelength
limit, and indeed $\tilde{\omega}_{2s}^{(p)}(q,k=0)$ and
$\tilde{\omega}_{1s}^{(p)}(q,k=0)$ are identical to the 
corresponding 3D and 2D plasmons (in Eq.~(5)) respectively with
$\tilde{n}_{3}=n_2/a$ and $\tilde{n}_2=n_1/a$. This is exactly what
one expects since the 
$D$-dimensional superlattice ``loses'' its discrete periodic structure
for $k=0$ and simply becomes the $(D+1)$-dimensional regular plasmon at
long wavelength.

Amazingly this does not happen for the Dirac plasma, where the
$D$-dimensional superlattice plasmon for $k=0$ does not become the
corresponding $(D+1)$-dimensional bulk plasma frequency as one expects
intuitively. In particular, $\tilde{\omega}_{2s}^{(l)}(q,k=0)$ would
agree with the 
corresponding 3D Dirac plasmon $\tilde{\omega}_3^{(l)}(q)$ (in
Eq.~(7)) only if we 
define the corresponding effective 3D density to be
\begin{equation}
\tilde{n}_3 = \left ( {9 \pi g}/{16} \right )^{1/4} \left ( {n_2}/{d^2}
\right )^{3/4},
\end{equation}
rather than the intuitive definition $\tilde{n}_3=n_2/d$. For the 1D
superlattice 
Dirac plasmon, the situation is qualitatively different since
$\tilde{\omega}_{1s}^{(l)}(q,k=0)$ does not depend at all on the carrier
density, and only 
the following strange substitution provides a correspondence between
$\tilde{\omega}_{1s}^{(l)}(q,k=0)$ and $\omega_{2}^{(l)}(q\rightarrow 0)$:
\begin{equation}
\tilde{n}_2 = g/(\pi d^2), 
\end{equation}
which is a constant for all carrier density. This strange lack of
correspondence between $\tilde{\omega}_{Ds}^{(l)}(k=0)$ and
$\omega_{D+1}^{(l)}$ is again a 
manifestation of the peculiar quantum nature of the Dirac plasma,
where the band-edge plasmon at $k=0$ in the $D$-dimensional
superlattice does not reduce to the corresponding bulk plasmon in
$(D+1)$-dimension, as it does for the ordinary parabolic ES.

In summary, we have found that the long-wavelength plasma frequency of
a massless Dirac plasma with linear carrier energy
dispersion is non-classical with an explicit $1/\sqrt{\hbar}$ appearing
in the plasma frequency. This is in sharp contrast with the widespread
expectation that the long-wavelength plasmon is a classical plasma
oscillation -- in fact, a massless Dirac plasma has 
no classical analogy. In addition, the long-wavelength
Dirac plasma frequency depends explicitly on the coupling constant
(``the fine structure constant'').
We have also shown that
the long wavelength plasma mode of a $D$-dimensional superlattice of
massless Dirac plasma does not reduce to the corresponding
$(D+1)$-dimensional bulk plasmon, as one would have expected
intuitively. All of these peculiar results follow from the fact that a
massless Dirac plasma is fundamentally non-classical since the energy
dispersion $E=vp$ characterizing a system with constant velocity (but
variable momentum) simply cannot happen in classical physics. We
believe that our predictions can be tested in doped graphene
layers and multilayers, and in doped graphene ribbons and multiribbons
arrays using electron scattering \cite{R1}, light scattering
\cite{R2}, or infrared \cite{R3} spectroscopies. But the real
importance of our results is conceptual as we 
establish a strange quantum behavior in the graphene world of a Dirac
plasma where the
long wavelength plasmon is explicitly non-classical in contrast to
plasmons in ordinary semiconductors and metals whose long wavelength
limit is necessarily a classical plasma frequency.

This work is supported by US-ONR, NSF-NRI, SWAN, and DOE-Sandia.

\end{document}